\documentclass[a4paper, twocolumn]{article}

\usepackage[T1]{fontenc}
\usepackage[utf8]{inputenc}
\usepackage[english]{babel}
\usepackage{amsmath}
\usepackage{amssymb}
\usepackage{cancel}
\usepackage{babel}
\usepackage{color}
\usepackage[colorlinks=true, allcolors=blue]{hyperref}
\usepackage{latexsym}
\usepackage{wasysym}
\usepackage{abstract}
\usepackage[a4paper,top=1.5cm,bottom=1.5cm,left=1.5cm,right=1.5cm,marginparwidth=1.75cm]{geometry}
\usepackage{balance}

\def\mn{_{\mu\nu}}	
\def\MN{^{\mu\nu}}
\def\lm{\mathcal{L}_m}
\def\R{{\mbox{\tiny (R)}}}
\def\RC{{\mbox{\tiny (C)}}}
\def\S{{\mbox{\tiny (S)}}}
\def\L{{\mbox{\tiny (${\cal L}$)}}}

\usepackage{xcolor}

\title{On Rastall gravity formulation as a $f(R,\lm)$ and a $f(R,T)$ theory}
\author{J\'ulio C. Fabris,$^{1,2,3}$ Oliver F. Piattella,$^{2,4}$ Davi C. Rodrigues$^{1,2}$}

\date{\textit{
$^1$ Departamento de F\'isica - Universidade Federal do Esp\'irito Santo, 29075-910 Vit\'oria, ES, Brazil\\ 
\vspace{0.2cm}
$^2$ N\'ucleo Cosmo-ufes - Universidade Federal do Esp\'irito Santo, 29075-910 Vit\'oria, ES, Brazil\\ 
\vspace{0.2cm}
$^3$ National Research Nuclear University MEPhI, Kashirskoe sh. 31, Moscow 115409, Russia\\
\vspace{0.2cm}
$^4$ Dipartimento di Scienza e Alta Tecnologia, Universit\`a degli Studi dell'Insubria e INFN, via Valleggio 11, I-22100 Como, Italy}}

\begin{document}

\twocolumn[
  \begin{@twocolumnfalse}
    \maketitle
    \begin{abstract}
Rastall introduced a stress-energy tensor whose divergence is proportional to the gradient of the Ricci scalar. This proposal leads to a change in the form of the field equations of General Relativity, but it preserves the number of degrees of freedom.  Rastall's field equations can be either interpreted as GR with a redefined SET, or it can imply different physical consequences inside the matter sector. We investigate limits under which the Rastall field equations can be directly derived from an action, in particular from two $f(R)$-gravity extensions: $f(R,\mathcal L_m)$ and $f(R,T)$.  We show that there are similarities between these theories, but the Rastall SET cannot be fully recovered from them, apart from certain particular cases here discussed. It is remarkable that a simple,  covariant and invertible redefinition of the SET, as the one proposed by Rastall, is hard to be directly implemented in the action. \\
\vspace{0.5cm}
    \end{abstract}
  \end{@twocolumnfalse}
]

\maketitle

\section{Introduction}
\label{Introduction}

In general relativity (GR), the stress-energy tensor (SET) is commonly defined as the variation of the matter action with respect to the metric. With this definition, and within GR, the SET conservation (i.e., $\nabla_\mu T^{\mu \nu} = 0$) is a consequence of both the field equations and diffeomorphism invariance \cite{Wald:1984rg}. On the other hand, for other theories of gravity, diffeomorphism invariance needs not to imply the SET conservation. This issue is detailed in Appendix \ref{sec:App}. 

Considering spacetime geometries beyond Minkowski, one could, in principle, suppose a generalization of the standard SET conservation relation as follows:
\begin{equation}\label{Rastallcons}
	\nabla_\mu  {T}^{ \R \mu\nu}{} \propto \nabla^\nu R\, .
\end{equation} 
This was the proposal of Rastall \cite{rastall}, and we use the superscript $^{\R}$ to specify a SET that satisfies the above relation. It should be pointed out that locally this SET conservation becomes $\partial_\mu T^{\R \mu \nu} = 0$ only if $\partial_\mu R = 0$ locally. In general, even though locally one can choose coordinates such that the metric is close to the Minkowski metric, this does not imply that the Ricci scalar (or its derivative) can be approximated by zero. Hence, Eq.~\eqref{Rastallcons} seems to define a relation that is incompatible with GR. That is, assuming that the SET is a quantity accessible by experiments, there seems to be a physical distinction between GR and the Rastall case.

On the other hand, Eq.~\eqref{Rastallcons} is not sufficient to define a theory. For instance, it could be used together with scalar-tensor gravity \cite{Fabris:2011wz}, or it could be implemented more closely to the GR context. In the latter case, the field equations read \cite{rastall, cosmo}
\begin{equation}
	\label{eq: Rastall Gmn = Tmn}
	G\mn = R\mn -\frac{1}{2} g\mn R = \kappa\left(T^\R_{\mu\nu} - \frac{\gamma - 1}{2}g_{\mu\nu}{T^\R}\right)\;,
\end{equation}
where $\kappa = 8\pi G$ and $\gamma$ is the free parameter of the theory (setting $\gamma = 1$ recovers GR with a standard SET). Using these field equations together with the Bianchi identities, the SET from Rastall's proposal satisfies
\begin{equation}
	\label{eq: Rastall conservation}
	\nabla^\mu{{T}^\R_{\mu\nu}} = \frac{\gamma - 1}{2}\nabla_\nu {T}^\R\;.
\end{equation}
Hence, it is also correct to see Rastall gravity as GR with a matter component whose SET satisfies Eq.~\eqref{eq: Rastall conservation} (see also Ref.~\cite{visser}). 

Related to this non-fundamental interpretation of Rastall gravity, some of us considered the case in which only one of the matter components has a SET of the Rastall type \eqref{eq: Rastall conservation} \cite{cosmo}.

From Eq.~\eqref{eq: Rastall conservation}, it is easy to see that the Rastall SET leads to a conserved rank-2 tensor $T^\RC_{\mu \nu}$, namely
\begin{align}
 	 & T^\RC_{\mu \nu} = \left (\delta^\sigma_\mu \delta^\kappa_\nu - \frac{\gamma - 1}2 g^{\sigma \kappa} g_{\mu  \nu} \right ) { T}^{\R}_{\sigma \kappa} \, , \label{TcalT} \\
 	 & \nabla^\mu T^\RC_{\mu \nu} = 0 \, .
\end{align}
Here, the superscript $^\RC$ is a reference to the conserved SET induced by the  Rastall approach. The 4-rank tensor in parenthesis can be inverted, implying that
\begin{equation} \label{eq: TR=LambdaTRC}
	T^\R\mn = \left( \delta^\sigma_\mu \delta^\kappa_\nu + \frac{\gamma -1}{2(3 - 2\gamma)} g^{\sigma \kappa}g_{\mu \nu} \right)T^\RC_{ \sigma \kappa} \, .
\end{equation}
This is the general relation between $T^\R\mn$ and $T^\RC\mn$, apart from a constant multiplicative factor and any divergence-free additive term.

Apart from the case $\gamma = 3/2$,\footnote{This case imposes that the Ricci scalar is always zero. We will not consider this case further.} it is always possible to change from $T^\RC_{\mu \nu}$ to ${ T}^\R_{\mu \nu}$ and back to $ T^\RC_{\mu \nu}$. Moreover, by expressing Eq.~\eqref{eq: Rastall Gmn = Tmn} through the use of $T^\RC\mn$, the Rastall field equations become identical to those of GR with the identification 
\begin{equation} \label{eq: TRC=TS}
	T^\RC\mn = T^\S\mn \, ,
\end{equation}
where $T^\S\mn$ is the standard SET defined from the matter action ($S_m$), that is:
\begin{equation}
	T^\S_{\mu \nu} \equiv -  \frac{2}{\sqrt{-g}} \frac{\delta S_m}{\delta g^{\mu \nu}}\,. \label{Tdeff}
\end{equation}

The above remarks show that all the information in the field equations of GR is also present in the field equations of Rastall theory, but written in a different way: it is a matter of properly converting from  $T^\RC\mn$ to $T^\R\mn$. Such conversion is always possible and it has an inverse. Therefore, in this sense of preserving the same information with respect to the SET, Rastall gravity and GR are equivalent (see also Ref.~\cite{visser}).

With respect to the action, there were attempts to define a consistent action that could straightforwardly derive the field equations \eqref{eq: Rastall Gmn = Tmn}. For instance, the action proposed by Smalley \cite{smalley} is devoted to find $T\mn^\R$ only and in principle it can do that, but it is not a scalar, hence it is not an action in the standard sense. We stress that the steps presented in this section do not provide an action capable of a straight derivation of $T\mn^\R$ either. However, following such steps, it is always possible to generate $T^\R\mn$ from a given standard matter action $S_m$. Indeed, from the latter one derives $T^\S\mn$, which is a conserved SET and hence it can be identified with $T^\RC\mn$. From eq.~\eqref{eq: TR=LambdaTRC} one finds $T^\R\mn$.

The inverse process of finding $S$ from a given $T\mn^\R$ is also possible and it is relevant since several applications of Rastall's theory start by assuming a given $T\mn^\R$. In general, from a given conserved SET it is expected that there should be an action, but finding the explicit action is not a trivial task. In Appendix \ref{sec:TRtoS} we illustrate this procedure.

The existence of such correspondence, nonetheless, does not imply that Rastall physics and GR physics are necessarily the same. The SET's $T^\RC_{\mu \nu}$ and ${ T}^\R_{\mu \nu}$ can be interpreted as suggesting different physical phenomena, a different way to match theoretical symbols to physical quantities. Such situation is far from being novel within different gravitational approaches. For instance, within scalar-tensor gravity, it is well known that there is a classical correspondence between the Jordan and Einstein frames (all the information in one is present in the other) \cite{Sotiriou:2007zu}. But, in spite of the formal equivalence, there are interpretations in which these frames do not describe the same physics \cite{GabrieleGionti:2020drq}. As another example, Palatini $f(R)$ gravity is identical to GR with a cosmological constant in vacuum \cite{Olmo:2011uz}. Since the latter two theories have the same number of degrees of freedom, inside matter one can in principle attribute any differences due to a SET redefinition (see for instance Ref.~\cite{Toniato:2019rrd} for an explicit construction). These examples are used here to stress that different SET expressions can be used (and are used) to motivate different physical approaches. Such physically different approaches are subject, in general, to experimental bounds. Thus, such bounds should be verified for the types of matter that are considered to be subject to the Rastall SET condition (see also Ref.~\cite{visser}).

About the equivalence between GR and Rastall's theory, note that in Ref.~\cite{darabi} conclusions are reached that are opposite to those of Ref.~\cite{visser}. In particular, phrased in our notation, $T^\RC\mn$ is really different from $T^\S\mn$ because the former contains intrinsically the curvature, via the field equations, whereas the latter does not. See also Refs.~\cite{Velten:2021xxw, Vanzella:2022xds} for a review and an investigation of the relation between Rastall's gravity and the dark components of our universe.

Our focus in this paper are certain non-standard actions that are claimed to be sufficient in order to find the Rastall equations. Apart from particular cases, we show otherwise. In the process, we develop a particular case that resembles the cosmology developed in Ref.~\cite{cosmo} with Rastall gravity.

In particular, we explore the differences and similarities between the field equations of Rastall and of two other gravitational theories that yield non-conserved SET's. More recently, $f(R)$ extensions of the type $f(R,\lm)$ \cite{harko1} and $f(R,T)$ were considered \cite{harko2}, where $\lm$ is the scalar part of the matter Lagrangian density and $T$ is the trace of the SET. These theories in general violate the standard SET conservation, but preserve diffeomorphism invariance. References \cite{brasil, nogales, Ziaie} consider the derivation of Rastall gravity from such theories. In these references, the authors propose a Rastall Lagrangian of the type $f(R,T) = R + \alpha T$. Specifically in Ref.~\cite{brasil} it is also claimed that Rastall gravity may be found using the following prescription $f(R, \lm) = \alpha R + \mathcal{G}(\lm)$, where $\mathcal{G}(\lm)$ is a general function of $\lm$.

In light of this, our main goal is to analyze if these models are consistent with Rastall gravity. We show that Rastall gravity cannot be obtained from a $f(R,T)$ or a $f(R,\lm)$ theory, as claimed. In order to achieve this goal, we  first analyze, in Sec.~\ref{sec: f(R lm)}, $f(R, \lm)$ theories and show how they cannot recover Rastall gravity; in the subsequent section, Sec. \ref{sec: f(R T)}, we explicitly calculate how $f(R,T)$ gravity behaves for several different choices of SETs, and demonstrate that it is possible to obtain a similar (but not identical) structure as that given by Rastall gravity but only for perfect fluids. On the other hand, even if $f(R,T)$ theories do not identically reproduce Rastall gravity, some of the similarities between the two classes of theories lead to some interesting correspondence in some specific cases. We fully develop a particular case in Sec. \ref{sec: LCDM}, and show a direct relation with the $\Lambda$CDM model of standard cosmology, but with a cosmological constant that becomes a dynamical dark energy component. We discuss our main results in Sec. \ref{sec: Discussion}.

\section{Similarities between $f(R, \lm)$ and Rastall theory}
\label{sec: f(R lm)}
\subsection{General case}

As an extension of $f(R)$ models \cite{f(R)}, a specific non-minimal coupling of the matter Lagrangian was proposed as a way to investigate non-geodesic motion of massive test particles \cite{Bertolami}. The complete generalization of this model, was named $f(R, \lm)$ theory and is described by the following action: 
\begin{equation}\label{fRLmaction}
    S = \kappa \int d^4x \sqrt{-g} f(R, \lm)\;.
\end{equation}
In general, it is not possible to split such action into a gravitational part plus a matter part. Nonetheless, one can define a SET as follows  \cite{Bertolami}:
\begin{equation} \label{eq: f(R,L) SET}
 	{T}\mn^{(\cal L)} \equiv -\frac{2}{\sqrt{-g}}\frac{\partial(\sqrt{-g}\lm )}{\partial g\MN} = g\mn\lm - 2\frac{\partial\lm}{\partial g\MN}\;. 
\end{equation}
The above SET definition is inspired by the standard definition \eqref{Tdeff}, but it does not require the existence of a matter action separated from the gravitational action. Although Eq.~\eqref{fRLmaction} is invariant under diffeomorphisms (i.e., it is a scalar action),  there is no guarantee  that ${T}\mn^{(\cal L)}$ is conserved.

Variation of the action \eqref{fRLmaction} yields:
\begin{equation}
	 \delta S \!=\! 
	 \kappa  \int d^4x \sqrt{-g}\left[f_R\delta R + f_{\lm}\delta\lm - \frac{1}{2}g_{\mu\nu}f\delta g^{\mu\nu}\right], 
\end{equation}
where we have defined $f_R \equiv \partial f/\partial R$ and $f_{\lm} \equiv \partial f/\partial \lm$. The variation  $\delta R$ is to be treated as usual. The variation of the matter Lagrangian due to $g\MN$ (denoted by $\delta_g{\lm}$) reads, using Eq.~\eqref{eq: f(R,L) SET}:
\begin{equation}
	\delta_g\lm = \frac{\partial \lm}{\partial g\MN} \delta g\MN = \frac{1}{2}\left(g_{\mu\nu}\lm - T^{(\cal L)}_{\mu\nu}\right)\delta g^{\mu\nu}\;.
\end{equation}
So, the field equations for the $f(R, \lm)$ theory are:
\begin{align}\label{fieldeqsfrlm}
    & f_R R\mn + (g\mn \Box - \nabla_\mu \nabla_\nu)f_R - \nonumber \\ 
    & \;\;  - \frac{1}{2} [f(R,\lm) - f_{\lm} \lm] g\mn     = \frac{1}{2} f_{\lm} T^{(\cal L)}\mn\;.
\end{align}
These field equations can be cast, using Eq.~\eqref{eq: f(R,L) SET}, as follows:
\begin{align}\label{fieldeqsfrlm2}
    G\mn = -\frac{f_{\lm}}{f_R}\frac{\partial\lm}{\partial g\MN} - \frac{1}{f_R}\mathcal{D}\mn f_R + \frac{1}{2}g\mn\left(\frac{f}{f_R} - R\right)\;,
\end{align}
where we have defined the operator:
\begin{equation}
    \mathcal{D}\mn \equiv (g\mn \Box - \nabla_\mu \nabla_\nu)\;.
\end{equation}
Taking the divergence of the field equations \eqref{fieldeqsfrlm2}, we find:
\begin{align}
    \nabla^\mu\left(-\frac{f_{\lm}}{f_R}\frac{\partial\lm}{\partial g\MN} - \frac{1}{f_R}\mathcal{D}\mn f_R + \frac{1}{2}g\mn\frac{f}{f_R}\right) - \frac{1}{2}\nabla_\nu R = 0\,,
\end{align}
and we may compare this with Eq.~\eqref{Rastallcons}. We could be tempted to conclude that:
\begin{equation}
    {T}^{\R}_{\mu\nu} = -\frac{f_{\lm}}{f_R}\frac{\partial\lm}{\partial g\MN} - \frac{1}{f_R}\mathcal{D}\mn f_R + \frac{1}{2}g\mn\frac{f}{f_R}\;,
\end{equation}
but the problem is that, in principle, $ {T}^{\R}_{\mu\nu}$ should not contain curvature terms. This issue can be addressed by taking the trace of the field equations, i.e.
\begin{align}
    -R = -\frac{f_{\lm}}{f_R}g\MN\frac{\partial\lm}{\partial g\MN} - \frac{3}{f_R}\Box f_R + 2\left(\frac{f}{f_R} - R\right)\;,
\end{align}
so that we can relate $R$ to the matter quantities.\footnote{This is a strategy adopted also in e.g. $f(R)$ gravity in the Palatini approach.} However, the above relation is not algebraic, due to the term $\Box f_R$. To this purpose, in order to kill the term $\mathcal{D}\mn f_R$, we need $f_R$ to be a constant, say $f_R = \alpha$. So $f$ must have the following form: 
\begin{equation}
	f(R, \lm) = \alpha R + \mathcal{G}(\lm)	\, , \label{fbrasil}
\end{equation}
where $\mathcal{G}(\lm)$ is a generic function of the matter Lagrangian. This Lagrangian was indeed considered in Ref.~\cite{brasil}, where the authors claim that Rastall gravity can be reproduced by it. 

The action corresponding to Eq.~\eqref{fbrasil} (apart from the constant $\alpha$) is exactly that of GR with $\lm \to {\cal G}(\lm)$. If one uses the standard SET definition \eqref{Tdeff} with 
\begin{equation}
	S_m = \kappa \int {\sqrt{-g}} d^4x\,\mathcal{G}(\lm)  \, ,
\end{equation} 
the derived SET will be conserved. One concludes that Eq.~\eqref{fbrasil} is just GR with the matter sector written in a different way. Qualitatively, this is what is necessary in order to find Rastall field equations.

To find a Rastall-like SET, the main point is to explore the differences between $T^\S\mn$ and $T^\L\mn$, as done in  Ref.~\cite{brasil}, but with a different approach. By computing $T^\S\mn$, its relation with $T^\L\mn$ becomes apparent, namely:
\begin{equation} \label{eq: TSfRL}
	T^\S_{\mu\nu} = \mathcal{G}' T^\L\mn + (\mathcal{G} - \mathcal{G}' \lm) g\mn\;,
\end{equation}
where $\mathcal{G} \equiv \partial \mathcal{G}/\partial \lm$.  As expected, for ${\cal G} = \lm$, we have $T^\S\mn = T^\L\mn$.

Since the split between the matter and the gravitational action  is trivial in this case, $T^\S\mn$ is necessarily conserved (as reviewed in Appendix \ref{sec:App}). Therefore, by identifying $T^\S\mn$ as $T^\RC\mn$ [as done in Eq.~\eqref{eq: TRC=TS}], one can use Eq.~\eqref{eq: TR=LambdaTRC} to find the  general Rastall SET that satisfies Eq.~\eqref{eq: Rastall conservation}, namely
\begin{align} \label{eq: TRfRL}
	T^\R\mn = & {\cal G}'\left ( T\mn^\L  + \frac{\gamma -1}{2(3-2 \gamma)} T^\L g\mn \right) + \nonumber \\[.1cm] 
	& + \frac{1}{3 - 2 \gamma} g\mn ({\cal G} - {\cal G}' \lm) \, .
\end{align}
It is straightforward to verify that Eq.~\eqref{eq: Rastall conservation} is satisfied for any ${\cal G}(\lm)$. The above expression is a generalization of Eq.~\eqref{eq: TR=LambdaTRC}: it becomes the latter for ${\cal G} = \lm$. 

The previous expression for $T^\R\mn$ \eqref{eq: TRfRL} was not shown in Ref.~\cite{brasil}, it is the general Rastall SET derived from a particular instance of $f(R,\lm)$ gravity \eqref{fbrasil}. Such solution does not really solve the issue of how to find a theoretical description in which the Rastall SET appears immediately or naturally. In essence, up to this point, we have transformed a conserved SET into a Rastall one (which can always be done).  Reference \cite{brasil} considers that $T^\L\mn$ is a Rastall-like (or Rastall-type) SET. As shown in Eqs.~(\ref{eq: TSfRL}, \ref{eq: TRfRL}), in general $T^\L\mn$ cannot be identified with $T^\R\mn$: there is only a superficial similarity since these SET's are not conserved. 

It would be curious if the Rastall SET could be identified with $T^\L\mn$, but it clearly cannot be identified as  such in general. In the following, we develop a particular case in which ${\cal G}' T^\L\mn$ is a Rastall SET. 

\subsection{$k$-essence with $T^\R\mn = {\cal G}'T^\L\mn$}

As an example, consider the Lagrangian of a free scalar field, i.e. 
\begin{equation}
\lm = -\frac 12 \nabla_\rho\phi\nabla^\rho\phi\, .
\end{equation}
For this case, the corresponding SET and its trace read:
\begin{align}
	& T^{(\cal L)}\mn = \nabla_\mu\phi\nabla_\nu\phi - \frac 12 g\mn \nabla_\rho\phi \nabla^\rho\phi\, , \\
	& T^{(\cal L)} = -\nabla_\rho\phi \nabla^\rho\phi\, .
\end{align}
This particular case is both simple and it has the property $\lm \propto T^\L$. Using the above, it is possible derive a Rastall SET more straightforwardly from $T^\L\mn$. Using the relations above, we can look for a procedure to obtain the Rastall SET structure from $T_{\mu\nu}^{\cal L}$. The previous expressions \eqref{TcalT}, \eqref{eq: TRC=TS} and \eqref{eq: TSfRL} suggest that the most direct way (and perhaps the only one) to 
obtain the Rastall gravity SET from ${\cal L}_m$ is by supposing:
\begin{equation} \label{eq:TRTLphiphi}
	T^\R\mn = {\cal G}'T^\L\mn \, . 
\end{equation}
For the above relation to be true, from Eq.~\eqref{eq: TRfRL} we find that 
\begin{equation}
	{\cal G}(\lm)=\lm {\cal G}'(\lm) (2 - \gamma),
\end{equation}
hence, 
\begin{equation}
	{\cal G} \propto \lm^{\frac{1}{2-\gamma}}  \propto \left(\nabla_\rho\phi \nabla^\rho\phi\right )^{\frac{1}{2-\gamma}} \, .	
\end{equation}
The case $\gamma = 2$  restricts the viable $\cal G$ functions to be null functions in order to make the identification \eqref{eq:TRTLphiphi} valid. It will not be considered.

In summary, we can rephrase our results as:
\begin{enumerate}
	\item Let $S$ be an action for  GR with $k$-essence,
		\begin{equation}
		    S = \int \left [ \frac{R}{2\kappa} -    \xi \left(\nabla_\rho\phi \nabla^\rho\phi\right)^\sigma \right] \sqrt{-g} d^4x \, ,
		\end{equation}
		where $\sigma$ and $\xi$ are constants.
	\item The field equations of this theory can be written in the form of Rastall field equations \eqref{eq: Rastall Gmn = Tmn}, with $2 - \gamma = 1/\sigma$ and
	\begin{equation}
	    T^\R\mn \propto   \left( \nabla_\mu\phi\nabla_\nu\phi - \frac 12 g\mn \nabla_\rho\phi \nabla^\rho\phi\right ) (\nabla_\rho\phi \nabla^\rho\phi)^{\sigma - 1} \, .
	\end{equation}
	\item Rastall approach has similarities to  a change from the standard SET $T^\S\mn$ to $T^\L\mn$. Nonetheless, these are only similarities, the Rastall SET cannot be truly derived in this way (even for the special case here considered with $\lm \propto T^\L$, which stresses further the similarities).
\end{enumerate}

\section{Similarities between $f(R,T)$ and Rastall theory}
\label{sec: f(R T)}

An approach which extends GR is to assume that the cosmological constant, responsible for the accelerated expansion of the universe, might be instead a dynamical term (see Ref. \cite{Lambdanotconstant} for a review in several different proposals). One possible formulation of this idea is to assume that the cosmological term depends on the trace of the SET, i.e. $T$. This is now dubbed as $\Lambda(T)$ gravity \cite{Lambda(T)}, and such model is an instance of the general class of $f(R,T)$ models.

On the other hand, as we have seen in the previous sections, there are in principle various definitions of the SET. So, which one enters $f(R,T)$ gravity? Since $T$ is a sort of independent field which modifies the action of gravity, it seems natural in this instance to suppose that matter is described by an independent action $S_m$, through which we are able to define a SET in the standard way, as in Eq.~\eqref{Tdeff}, compute its trace and use it into the function $f(R,T)$.

So, we consider the general $f(R,T)$ Lagrangian as given by
\begin{equation}
\label{fRTlag}
{\cal L} = \frac{1}{2\kappa} \sqrt{-g} f(R,T) + \sqrt{-g}\lm\;.
\end{equation}
Using the standard definition of the SET as in Eq.~\eqref{Tdeff}, the field equations are the following \cite{harko2}:
\begin{eqnarray}\label{fieldeqsfRT}
	f_R(R,T)R_{\mu\nu} - \frac{1}{2}f(R,T)g_{\mu\nu}\nonumber\\ + (g_{\mu\nu}\Box - \nabla_\mu\nabla_\nu)f_R(R,T) \nonumber\\
=  \kappa T_{\mu\nu}
- f_T(R,T)(T_{\mu\nu} + \Theta_{\mu\nu})\;,
\end{eqnarray}
where,
\begin{equation} \label{eq: Theta definition}
	\Theta\mn \equiv g^{\rho\sigma}\frac{\partial T_{\rho\sigma}}{\partial g^{\mu\nu}}\;,
\end{equation}
and we omit the superscript $^\S$ for the standard SET in this section.

As in the previous section, we investigate whether a $f(R,T)$ theory is able to reproduce Rastall gravity, i.e. Eq.~\eqref{eq: Rastall Gmn = Tmn}. To this purpose, we make again a simplifying ansatz for the functional form of $f(R,T)$, i.e. we choose 
\begin{equation}
\label{escolha}
	f(R,T) = \alpha R + \kappa\beta T\;,
\end{equation}
with $\alpha$ and $\beta$ arbitrary parameters. The field equations \eqref{fieldeqsfRT} become then:
\begin{equation}\label{fieldeqsfRTchoice}
	G\mn = \kappa\left[\frac{ 1 - \beta}{\alpha}T_{\mu\nu} + \frac{\beta}{2\alpha}\left(g_{\mu\nu}T - 2\Theta_{\mu\nu}\right)\right]\;,
\end{equation}
so that the SET conservation, for $\beta \not=1$, is written as
\begin{equation}
	\nabla_\mu{T^{\mu\nu}} = \frac{\beta}{2(\beta - 1)}\left(\nabla^\nu T  - 2{\nabla_\mu\Theta^{\mu\nu}}\right)\;,
\end{equation}
whereas for $\beta = 1$ one has:
\begin{equation}
	\nabla_\mu{\Theta^{\mu\nu}} = \frac{1}{2}\nabla^\nu T\;.
\end{equation}
As one can see, for $\beta\neq 1$ Rastall theory can be obtained (i.e. the standard definition of the SET is equivalent the one of the Rastall SET) if $\nabla_\mu\Theta^{\mu\nu} = 0$ and if the following relations between the parameters $\alpha$ and $\beta$ are satisfied:
\begin{equation}
	\alpha = 1 - \beta\;, \qquad \beta = \frac{\gamma - 1}{\gamma - 2}\;.
\end{equation}
Let us consider now the explicit form of $\Theta_{\mu\nu}$ for some relevant cases. In the subsequent calculations we use the following relations:
\begin{eqnarray}\label{relsmetric}
	\frac{\partial g^{\alpha \beta}}{\partial g^{\mu\nu}} = \delta^\alpha_\mu\delta^\beta_\nu\;,\quad
\frac{\partial g_{\alpha\beta}}{\partial g^{\mu\nu}} =   - g_{\alpha\mu}g_{\beta\nu}\;.
\end{eqnarray}
Since:
\begin{equation}
    T\mn = g\mn\lm - 2\frac{\partial\lm}{\partial g\MN}\;,
\end{equation}
then from Eq.~\eqref{eq: Theta definition} we have:
\begin{equation}\label{Thetamnformula}
    \Theta\mn = -2T\mn + g\mn\lm - 2g^{\rho\sigma}\frac{\partial^2\lm}{\partial g^{\rho\sigma}g\MN}\;.
\end{equation}
See Ref.~\cite{harko2} for more details.

\subsection{The electromagnetic case}

For the Lagrangian $\lm = - F_{\mu\nu}F^{\mu\nu}/4$ it is straightforward to compute the standard SET, which reads:
\begin{equation}
    T\mn = F_{\mu\alpha}F_{\nu\beta}g^{\alpha\beta} + g\mn\lm\;,
\end{equation}
and from Eq.~\eqref{eq: Theta definition}, using the relations \eqref{relsmetric}, it is straightforward to obtain:
\begin{equation}
	\Theta_{\mu\nu} = - T_{\mu\nu}\;.
\end{equation}
Using this formula, and the fact that the electromagnetic SET is traceless, the field equations, Eq. \eqref{fieldeqsfRT} become:
\begin{equation}
	G\mn = \frac{\kappa}{\alpha}T_{\mu\nu}\;,
\end{equation}
and thus, the conservation of the SET is
\begin{equation}
	\nabla_\mu{T^{\mu\nu}} = 0\;.
\end{equation}
That is, this is GR ($\alpha$ can be incorporated in the electromagnetic field or in $\kappa$). This result had to be expected since the beginning, due to the tracelessness of the electromagnetic SET.

Therefore, until now we have found no surprising result, since for the same case Rastall gravity also reduces to GR. Indeed, since for the electromagnetic field $T = 0$, one has $f(R,T) = f(R,0) = f(R)$. So, the theory reduces to a $f(R)$ theory.

\subsection{Perfect-fluid case}

The SET of a perfect-fluid is
\begin{equation} \label{eq: perfect fluid SET}
	T_{\mu\nu} = (\rho + p)u_\mu u_\nu -p g\mn\;.
\end{equation}
If for the perfect fluid we adopt the Lagrangian $\lm = -p$, then from Eq. \eqref{Thetamnformula}, we have:
\begin{equation}
\label{fluido1}
\Theta_{\mu\nu} = - 2T_{\mu\nu} - pg_{\mu\nu}\;.
\end{equation}
With this expression, the field equations \eqref{fieldeqsfRTchoice} read:
\begin{align}\label{perfectfluidfieldeqs}
	G\mn &= \kappa\left[\frac{1 + \beta}{\alpha} T_{\mu\nu} + \frac{\beta}{2\alpha}g_{\mu\nu}(T + 2p)\right]\;.
\end{align}
For the special case $\beta=-1$, the field equations and the conservation of the SET are
\begin{align}\label{perfectfluidfieldeqsbetaminus1}
	G\mn &=- \frac{4\pi G}{\alpha} g_{\mu\nu}(T + 2p)\;.
\end{align}
The divergence of this equation gives:
\begin{equation}
	\nabla_\mu (T + 2p) = 0\;.
\end{equation}
So, in this case, only the trace of the SET satisfies a conservation equation, which, being $T = \rho - 3p$, can be integrated to give:
\begin{align}\label{equationofstate}
	p = \rho + \Lambda\;,
\end{align}
where $\Lambda$, usually referred to as the cosmological constant, now appears as an integration constant. In general, for $\Lambda \neq 0$, the equation of state becomes somehow a mixture of a stiff matter fluid and a cosmological constant. 

For $\beta \not = -1$, the modified Einstein equations are neither identical to GR or Rastall. 

Recently, in Ref.~\cite{Haghani:2023uad} a new prescription is proposed, instead of Eq.~\eqref{fluido1}:
\begin{equation}
    \Theta_{\mu\nu} = -3T_{\mu\nu} - \frac{1}{4}(7p - \rho)g_{\mu\nu}\,.
\end{equation}
So that Eq.~\eqref{fieldeqsfRTchoice} becomes, instead of Eq.~\eqref{perfectfluidfieldeqs}:
\begin{equation}\label{perfectfluidfieldeqs2}
	G\mn = \kappa\left[\frac{1 + 2\beta}{\alpha}T_{\mu\nu} + \frac{\beta}{4\alpha}g_{\mu\nu}\left(\rho + p\right)\right]\;,
\end{equation}
For the case $\beta = -1/2$, Bianchi identity implies here:
\begin{equation}
    p = -\rho + \Lambda\,,
\end{equation}
that is the equation of state of vacuum energy plus, again, a cosmological constant.

\subsection{Scalar field case}

The Lagrangian of a self-interacting scalar field $\phi$, subject to a generic potential $V(\phi)$ is 
\begin{equation}
\label{escalar1}
	\lm = - \frac{1}{2}\nabla^\rho\phi\nabla_\rho\phi + V(\phi)\;,
\end{equation}
from which the standard SET, using Eq.~\eqref{Tdeff}, is:
\begin{equation}\label{phiTmunu}
    T\mn = \nabla_\mu\phi\nabla_\nu\phi + g\mn\lm\,,
\end{equation}
and $\Theta\mn$ is\footnote{In Ref. \cite{harko2} a factor 2 is missing multiplying the first term of the right-hand-side, and it is probably a misprint. This factor appears in another approach, through the fluid representation, as we will see later. However, it does not change the main aspect of the analysis for our present purposes.} 
\begin{equation}
\label{phiThetamunu}
	\Theta\mn = - 2T_{\mu\nu} + \frac{1}{2}g_{\mu\nu}T - g_{\mu\nu}V\;.
\end{equation}
The field equations, for the ansatz \eqref{escolha}, are then given by
\begin{align}
	G\mn &= \frac{8 \pi G}{\alpha} [ (1+\beta) T\mn + \beta g\mn V ]\;,
\end{align}
and the equation of motion for the scalar field is, in general:
\begin{equation}
    \left(\frac{f_T}{\kappa} + 2\right)\Box\phi + \left(\frac{2f_T}{\kappa} + 1\right)V_\phi = 0\,,
\end{equation}
where $V_\phi$ represents the derivative of $V$ with respect to $\phi$. For our ansatz \eqref{escolha}, becomes:
\begin{equation}
    (\beta + 1)\Box\phi + (2\beta + 1)V_\phi = 0\,.
\end{equation}
For the special case $\beta = -1$, the field equations become:
\begin{align}
	G\mn &= -\frac{\kappa}{\alpha} g\mn V \;, 
\end{align}
and the potential is a constant. Moreover, curiously, there is no dynamics for the scalar field, since its equation of motion reduces to $V_\phi = 0$, which means that we have to choose the scalar field which makes extremal the potential $V$ (it could be a local maximum or minimum). If no extremal points exist, then there are no solutions for the special case $\beta = -1$. A constant potential acts then as a cosmological constant.

For $\beta\neq -1$, the two field equations are:
\begin{align}
	G\mn &= \frac{8 \pi G}{\alpha} \Big[(1+\beta) \left(\nabla_\mu\phi\nabla_\nu \phi - \frac{1}{2} g\mn \nonumber \nabla_\rho\phi\nabla^\rho\phi  \right)  \\
	& \hspace{4cm}+ (1+2\beta) g\mn V  \big] \;, \\
	\Box \phi &= - \frac{1+ 2 \beta}{1+\beta} V_\phi\;.
\end{align}
One can then see that GR can be recovered by a trivial redefinition of the scalar field and the potential as follows:
\begin{eqnarray}
\sqrt{\frac{1 + \beta}{\alpha}}\phi \longrightarrow \phi\;, \qquad \frac{1+ 2 \beta}{\alpha} V \longrightarrow V\;.
\end{eqnarray}
One aspect, however, must be remarked: depending on the sign of $(1 + \beta)/\alpha$, an ordinary scalar field can become phantom and an attractive potential can become repulsive. Even so, the general structure is not the same as found in the corresponding case of Rastall gravity.

\section{Similarities between Rastall and $f(R,T)$ cosmology}
\label{sec: LCDM}

We have seen that Rastall theory cannot be framed in a $f (R, T )$ theory. Nonetheless, the two theories share a similar cosmological behavior. We will give below an example of this similarity in a specific case of a fluid that is split into two interacting components, one representing dark matter, the other representing dark energy.

In the first subsection we address the cosmic expansion and then we subsequently work out the evolution of small perturbations.

\subsection{Background evolution - case 1}

Let us set $\alpha = 1 + \beta$ in Eq.~\eqref{perfectfluidfieldeqs}. This redefinition is possible since it amounts to multiply the Lagrangian by a global constant factor. Of course, in what follows $\beta \neq -1$. Therefore, we have:
\begin{align}
	G\mn &= \kappa\left[T_{\mu\nu} + \frac{\beta}{2(1 + \beta)}g_{\mu\nu}(T + 2p)\right]\;,\\
	\nabla_\mu{T^{\mu\nu}} &= - \frac{\beta}{2(1 + \beta)}\nabla^\nu\left(T + 2p\right)\;.
\end{align}
Now, let us consider a spatially flat cosmic metric,
\begin{equation}
ds^2 = dt^2 - a(t)^2\delta_{ij}dx^idx^j\;,
\end{equation}
and a perfect fluid Lagrangian, Eq. \eqref{eq: perfect fluid SET},
with $\rho$ and $p$ depending only on the time coordinate. The modified Friedmann equations are:
\begin{align}
3\left(\frac{\dot a}{a}\right)^2 = \kappa\left[\frac{2 + 3\beta}{2(1 + \beta)}\rho - \frac{\beta}{2(1 + \beta)}p\right]\;,\\
\left[\frac{2 + 3\beta}{2(1 + \beta)}\right]\dot\rho - \frac{\beta}{2(1 + \beta)}\dot p + 3H(\rho + p) = 0\;,
\end{align}
where the dot denoted derivative with respect to the cosmic time.

We decompose the fluid into two components, a pressureless matter $\rho_m$ ($p_m = 0$) and a cosmological term $\rho_\Lambda$ ($p_\Lambda = - \rho_\Lambda$), such that,
\begin{align}
\rho = \rho_m + \rho_\Lambda\;, && p = p_m + p_\Lambda = - \rho_\Lambda\;.
\end{align}
Then, both equations can be rewritten as
\begin{align}
3\left(\frac{\dot a}{a}\right)^2 = \kappa\left[\frac{2 + 3\beta}{2(1 + \beta)}\rho_m + \frac{1 + 2\beta}{1 + \beta}\rho_\Lambda\right]\;,\\
\left[\frac{2 + 3\beta}{2(1 + \beta)}\right]\dot\rho_m + \frac{1 + 2\beta}{1 + \beta}\dot\rho_\Lambda + 3H\rho_m = 0\;.
\end{align}
Since we want the cosmic history and also structure formation to be preserved, we assume that the matter component conserves separately. In order to preserve consistency, the two fluids need to be interacting. This is not an issue, since our fluid is actually only one and the decomposition made is thus fictitious. The conservation of the matter part of the fluid is then the usual one:
\begin{equation}
	\dot\rho_m + 3H\rho_m = 0\;,
\end{equation}
and, therefore,
\begin{equation}
	\rho_m = \frac{\rho_{m0}}{a^3}\;,
\end{equation}
in which we have defined the integration constant $\rho_{m0}$ as the matter density at present time. With this hypothesis, the previous two equations become:
\begin{align}
3\left(\frac{\dot a}{a}\right)^2 = \kappa\left[\frac{2 + 3\beta}{2(1 + \beta)}\rho_m + \frac{1 + 2\beta}{1 + \beta}\rho_\Lambda\right]\;,\\
\left[\frac{\beta}{2(1 + \beta)}\right]\dot\rho_m + \frac{1 + 2\beta}{1 + \beta}\dot\rho_\Lambda = 0\;. \label{eq: continuity after rhom} 
\end{align}
Then, Eq.~\eqref{eq: continuity after rhom} can be integrated, leading to
\begin{equation}
\rho_\Lambda = - \frac{\beta}{2(1 + 2\beta)}\rho_m + \rho_{\Lambda0}\;,
\end{equation} 
where $\rho_{\Lambda0}$ is the cosmological constant density at present time. In consequence of this, we finally get to
\begin{align}
3\left(\frac{\dot a}{a}\right)^2 = \kappa(\rho_m + \bar\rho_{\Lambda0})\;,\\
\bar\rho_{\Lambda0} = \frac{1 + 2\beta}{1 + \beta}\rho_{\Lambda0}\;.
\end{align}
Hence, the standard $\Lambda$CDM model is reproduced. All the background tests are thus equally satisfied, as explained in Ref.~\cite{cosmo}.

We have been able to show that from a intrinsically non-conservative modified theory of gravity, $f(R,T)$, the $\Lambda$CDM model is re-obtained. On the other hand, the cosmological constant does not behave as in GR, since the non-conservation of this component changes its evolution.

\subsection{Background evolution - case 2}

Let us set $\alpha = 1 + 2\beta$ in Eq.~\eqref{perfectfluidfieldeqs2}. In what follows $\beta \neq -1/2$. We have:
\begin{align}
	G\mn &= \kappa\left[T_{\mu\nu} + \frac{\beta}{2(1 + 2\beta)}g_{\mu\nu}(\rho + p)\right]\;,\\
	\nabla_\mu{T^{\mu\nu}} &= - \frac{\beta}{2(1 + 2\beta)}\nabla^\nu\left(\rho + p\right)\;.
\end{align}
The modified Friedmann equations are:
\begin{align}
3\left(\frac{\dot a}{a}\right)^2 = \kappa\left[\frac{2 + 5\beta}{2(1 + 2\beta)}\rho + \frac{\beta}{2(1 + 2\beta)}p\right]\;,\\
\left[\frac{2 + 5\beta}{2(1 + 2\beta)}\right]\dot\rho + \frac{\beta}{2(1 + 2\beta)}\dot p + 3H(\rho + p) = 0\;,
\end{align}
Now, if we decompose the fluid into two components as we did earlier, both equations can be rewritten as:
\begin{align}
3\left(\frac{\dot a}{a}\right)^2 = \kappa\left[\frac{2 + 5\beta}{2(1 + 2\beta)}\rho_m + \rho_\Lambda\right]\;,\\
\left[\frac{2 + 5\beta}{2(1 + 2\beta)}\right]\dot\rho_m + \dot\rho_\Lambda + 3H\rho_m = 0\;.
\end{align}
Again assuming:
\begin{equation}
	\dot\rho_m + 3H\rho_m = 0\;,
\end{equation}
the previous two equations become:
\begin{align}
3\left(\frac{\dot a}{a}\right)^2 = \kappa\left[\frac{2 + 5\beta}{2(1 + 2\beta)}\rho_m + \rho_\Lambda\right]\;,\\
\left[\frac{\beta}{2(1 + 2\beta)}\right]\dot\rho_m + \dot\rho_\Lambda = 0\;. \label{eq: continuity after rhom2} 
\end{align}
Equation~\eqref{eq: continuity after rhom2} can be integrated, leading to the same result obtained earlier:
\begin{equation}
\rho_\Lambda = - \frac{\beta}{2(1 + 2\beta)}\rho_m + \rho_{\Lambda0}\;,
\end{equation} 
from which:
\begin{align}
3\left(\frac{\dot a}{a}\right)^2 = \kappa(\rho_m + \rho_{\Lambda0})\;.
\end{align}
So, for the fluid splitting chosen here, the new proposal of Ref.~\cite{Haghani:2023uad} does not change the final result. One simply avoids the need of defining $\bar\rho_{\Lambda0}$. 

\subsection{Evolution of small matter perturbations - case 1}

So far, we obtained that the $\Lambda$CDM model for the background evolution is recovered. We now show that this is also the case for small perturbations. In order to carry out this investigation, let us write the equations in the alternative form,
\begin{align}
	R_{\mu\nu} &= \kappa\left[T_{\mu\nu} - \frac{(1 + 2\beta)}{2(1 + \beta)}g_{\mu\nu}T - \frac{\beta}{1 + \beta}g_{\mu\nu}p\right]\;,\\
	\nabla_\mu{T^{\mu\nu}} &= - \frac{\beta}{2(1 + \beta)}\nabla^\nu\left(T + 2 p\right)\;. \label{eq: SET perturbation general}
\end{align}
We choose a linearly perturbed FLRW metric, with the perturbation variable denoted as $h_{\mu\nu}$, in the synchronous coordinate condition, i.e. $h_{\mu0} = 0$ \cite{ma}. Our field equations are covariant and we have not assumed diffeomrophism invariance to be broken, so this choice is possible. Then, we define,
\begin{equation}
h = \frac{h_{kk}}{a^2}\;,
\end{equation}
and similarly to what we have done in the previous section, we split the SET into
\begin{equation}
T^{\mu\nu} = T^{\mu\nu}_m + T^{\mu\nu}_\Lambda\;,
\end{equation}
with,
\begin{align}
T^{\mu\nu}_m = \rho_m u^\mu u^\nu\;, && T^{\mu\nu}_\Lambda = \rho_\Lambda g^{\mu\nu}\;.
\end{align}
Once again, applying the same reasoning as before, we assume that the matter SET conserves separately
\begin{equation}
\nabla_\mu{T^{\mu\nu}} = 0\;.
\end{equation}
Hence, we can rewrite Eq. \eqref{eq: SET perturbation general} as:
\begin{equation}
\nabla_\mu{T^{\mu\nu}_\Lambda} = - \frac{\beta}{2(1 + \beta)}\nabla^\nu\left(T + 2 p\right)\;.
\end{equation}
We follow closely the perturbative analysis presented in Ref. \cite{cosmo}. The set of the perturbed equations are then:
\begin{align}
	\frac{\ddot h}{2} + H\dot h &= \kappa\left(\frac{1}{2 (1 + \beta)}\delta\rho_m - \frac{1 + 2\beta}{1 + \beta}\delta\rho_\Lambda\right)\;,\\
\dot\delta_m &= \frac{\dot h}{2}\;, \\
\delta\dot\rho_\Lambda &= - \frac{\beta}{2(1 + 2\beta)}\delta\dot\rho_m\;, \label{eq:deltarhomdeltaLambda}
\end{align}
with the usual definition for the density contrast for the matter component as:
\begin{equation}
\delta_m = \frac{\delta\rho_m}{\rho_m}\;.
\end{equation}
Note that the perturbations of the four velocities of the cosmological term and of the pressureless matter are zero. The first is a choice, allowed from the residual gauge freedom typical of the synchronous gauge, the second, for the cosmological term, is a direct consequence of the equation of state $p_\Lambda = - \rho_\Lambda$. 

Combining these equations, we obtain:
\begin{equation}
\ddot\delta_m + 2H\dot\delta_m - 4\pi G\rho_m \delta_m = 0\;.
\end{equation}
Therefore, the evolution of small matter fluctuations in the model under consideration is also identical to the one in the $\Lambda$CDM model of GR. Despite the similarities, an important difference is that the cosmological constant introduced in the $f(R,T)$ theory is not really a constant, but has an evolution dictated by the non-conservative character of the theory. This feature is the same as it was found in Ref. \cite{cosmo} for the standard Rastall gravity.

In some sense, it is not surprising that the $\Lambda$CDM solution is recovered also at the perturbative level, for the following reason. Since we have chosen $p_m = 0$ and $p_\Lambda = -\rho_\Lambda$, pressure is not an extra degree of freedom. Therefore, from Eq.~\eqref{eq: SET perturbation general} it is clear that we obtain an equation in which a linear combination of the derivatives of the energy densities is vanishing, cf. Eq.~\eqref{eq: continuity after rhom} and Eq.~\eqref{eq:deltarhomdeltaLambda}. Therefore, the two energy densities are equal up to a constant and since we have imposed $\rho_m$ to evolve as ordinary matter, we then recover the $\Lambda$CDM model. Actually this argument is valid non-perturbatively, since it can be applied directly to Eq.~\eqref{eq: SET perturbation general}, from which we get:
\begin{equation}
	\nabla^\nu\left[(1 + 2\beta)\rho_\Lambda + \beta\rho_m\right] = 0\;,
\end{equation}
provided again that the matter SET conserves separately.

\subsection{Evolution of small matter perturbations - case 2}

Let us write the equations using the prescription of Ref.~\cite{Haghani:2023uad}:
\begin{align}
	R_{\mu\nu} &= \kappa\left[T_{\mu\nu} - \frac{1 + 3\beta}{2(1 + 2\beta)}g_{\mu\nu}\rho + \frac{3 + 5\beta}{2(1 + 2\beta)}g_{\mu\nu}p\right]\;,\\
	\nabla_\mu{T^{\mu\nu}} &= - \frac{\beta}{2(1 + 2\beta)}\nabla^\nu\left(\rho +  p\right)\;. \label{eq: SET perturbation general2}
\end{align}
Again, we split the SET into
\begin{equation}
T^{\mu\nu} = T^{\mu\nu}_m + T^{\mu\nu}_\Lambda\;,
\end{equation}
and we assume that the matter SET conserves separately, so we can rewrite Eq. \eqref{eq: SET perturbation general2} as:
\begin{equation}
\nabla_\mu{T^{\mu\nu}_\Lambda} = - \frac{\beta}{2(1 + 2\beta)}\nabla^\nu\left(\rho +  p\right)\;.
\end{equation}
The set of the perturbed equations are then:
\begin{align}
	\frac{\ddot h}{2} + H\dot h &= \kappa\left[\frac{1 + \beta}{2 (1 + 2\beta)}\delta\rho_m - \delta\rho_\Lambda\right]\;,\\
\dot\delta_m &= \frac{\dot h}{2}\;, \\
\delta\dot\rho_\Lambda &= - \frac{\beta}{2(1 + 2\beta)}\delta\dot\rho_m\;. \label{eq:deltarhomdeltaLambda}
\end{align}
Combining these equations, we obtain:
\begin{equation}
\ddot\delta_m + 2H\dot\delta_m - 4\pi G\rho_m \delta_m = 0\;,
\end{equation}
as we have found earlier. So, the two different prescriptions do not change the results concerning the cosmological model discussed here.

\section{Discussion and conclusions}
\label{sec: Discussion}
 
In this paper we analyzed the possibility of conceiving a Lagrangian for Rastall theory. First, we showed that theories of type $f(R, \lm)$ cannot reproduce Rastall gravity. Even assuming $f(R, \lm) = \alpha R + \mathcal{G}(\lm)$, this structure either recovers GR, or strongly deviates from Rastall gravity. We have also showed that, in spite of the claims in the literature, theories of the class $f(R,T)$ cannot either reproduce Rastall gravity: the only way to cast Rastall theory in the framework of $f(R,T)$ gravity is through the hydrodynamical representation of a scalar field. Hence, the original proposal of Rastall gravity seems to resist to attempts of formulating it in a Lagrangian formalism.

We remark that, in spite of the formal general differences addressed in Section \ref{sec: f(R T)}, in section \ref{sec: LCDM} we could reproduce, in the context of $f(R,T)$ gravity, one cosmological picture of Rastall gravity that has received considerable attention \cite{cosmo}.

We have also stressed the differences and similarities between Rastall, $f(R,T)$ and $f(R,\lm)$ formulations. We note that all these theories have a SET that is not conserved. However, this does not imply that they violate diffeomorphism invariance (as explained in Appendix \ref{sec:App}) nor that it is impossible to recast them in a new form in which another SET is conserved. We hope that  the relations here uncovered, and the formulations provided by these theories, which are commonly considered in the literature \cite{brasil, nogales, Ziaie}, may clarify their interrelationship and be helpful on the understanding of gravity and cosmology.

\bigskip

\noindent
{\bf Acknowledgements:} The authors are indebted with Jos\'e A.C. Nogales for fruitful discussions about the subject of this work. The authors thank FAPES (Brazil) and CNPq (Brazil) for partial support. 

\appendix

\setcounter{equation}{0}
\renewcommand{\theequation}{\thesection\arabic{equation}}

\section{Diffeomorphism invariance and the conservation of $T_{\mu \nu}$}\label{sec:App}

Here we review the demonstration and the necessary conditions about the relation between $ \nabla^\mu T_{\mu \nu}=0$ and diffeomorphism invariance. The demonstration here presented is in part based on Ref.~\cite{Wald:1984rg}. The relevant assumptions for the context of this work are emphasized.

Consider an action $S[g,\Psi]$ which is a functional of the spacetime metric $g_{\mu \nu}$ and additional fields which are collectively denoted by $\Psi$. The fields denoted by $\Psi$ can be of any tensorial nature (i.e., scalars, vectors, second-rank tensors ...). Under a change of spacetime coordinates, or, more precisely, a diffeomorphism between the spacetime manifold into itself, the action $S$ must be invariant, since it is a scalar. On the other hand, the components of the fields $g_{\mu \nu}$ and $\Psi$ need to transform covariantly with respect to the change of coordinates. Namely, consider a change of coordinates generated by $\xi^\mu$ and given by
\begin{equation}
	x'^\mu = x^\mu + \xi^\mu(x) \, . \label{diffXi}
\end{equation}
The change above induces the following change in the metric components,
\begin{equation}
	\delta_\xi g_{\mu \nu} = {\cal L}_{\xi} g_{\mu \nu} = \nabla_\mu \xi_\nu + \nabla_\nu \xi_\mu \, , \label{diffg}
\end{equation}
where $\cal L$ is the Lie derivative.

In the context of general modified gravity theories, it is not always clear how to split the complete action into a matter and a gravitational part, sometimes there may be no natural split or it may be impossible. Nonetheless, as a first step, let us assume that one can write the following split
\begin{equation}
	S[g,\Psi] = S_G[g,\phi] + S_m[g, \psi] \, , \label{splitGM}
\end{equation}
where $S_G$ and $S_m$ refer, respectively, to the gravitational and the matter parts; $\Psi$ was decomposed into the gravitational fields besides the metric, denoted by $\phi$, and the matter fields, denoted by $\psi$. As an example of a theory that is commonly decomposed as above, we recall the Brans-Dicke theory (in the Jordan frame), where the Brans-Dicke scalar field $\phi_{\rm BD}$ is commonly interpreted as a gravitational field, and it does not (in the Jordan frame) appear inside $S_m$. 

Let the SET be defined as in Eq.~\eqref{Tdeff}. Since in this appendix this is the single SET used, we omit the superscript $^\S$. Since $S_m$ is a scalar, one can write that, under the coordinates change given by Eq.~\eqref{diffXi},
\begin{align}
	0 &= \delta_\xi S_m  \label{varXiSM} \\
	& = \int \frac{\delta S_m}{\delta g^{\mu \nu}(x)} \delta_\xi g^{\mu \nu}(x) \, d^4 x + \int \frac{\delta S_m}{\delta \psi(x)} \delta_\xi \psi(x) \, d^4 x \, .\nonumber
\end{align}
Since $\psi$ does not appear in $S_G$, the complete equations of motion for $\psi$ are given by $\delta S_m / \delta \psi =0$. Therefore, using the matter field equations and Eqs.~(\ref{diffg}, \ref{Tdeff}, \ref{varXiSM}),
\begin{equation}
	0 = \int T_{\mu \nu} \nabla^\mu \xi^\nu \sqrt{-g} \, d^4x = -\int \nabla^\mu T_{\mu \nu}  \xi^\nu \sqrt{-g} \, d^4x \, .
\end{equation}
The relation above needs to hold for any $\xi^\nu$, therefore
\begin{equation}
	\nabla^\mu T_{\mu \nu} = 0\, .
\end{equation}
We conclude that invariance of the action $S_m$ under the diffeomorphism \eqref{diffXi} leads to the conservation of $T_{\mu \nu}$. Equivalently, one can also state, in this context, that  $T_{\mu \nu}$ is the Noether current induced by the $S_m$ invariance under diffeomorphism. 

On the other hand, care should be taken on the assumption that any $T_{\mu \nu}$ must be conserved due to diffeomorphism invariance of the action. It should be clear that the split of $S$ between $S_G$ and $S_m$, together with the definition \eqref{Tdeff}, are not innocuous assumptions. For instance, consider a Brans-Dicke action with the nonstandard assumption that its scalar field $\phi_{\rm BD}$ should be considered as a matter field (e.g., the scalar field kinetic term would be considered as part of the $S_m$ action). In this case, the split \eqref{splitGM} could be written as 
\begin{equation}
	S = S_G[g, \phi_{\rm BD}] +  S_m[g, \phi_{\rm BD}, \psi]	 \, .
\end{equation}
Using the above split and Eq.~\eqref{Tdeff}, it should be clear that $T_{\mu \nu}$ needs not be conserved, even though the action $S$ and its parts $S_G$ and $S_m$ are scalars. In this case, this $T_{\mu \nu}$ is not the Noether current associated with diffeomorphism invariance. In fact, we have:
\begin{align}
	0 &= \delta_\xi S_m   \\
	& = \int \frac{\delta S_m}{\delta g^{\mu \nu}(x)} \delta_\xi g^{\mu \nu}(x) \, d^4 x + \int \frac{\delta S_m}{\delta \phi_{\rm BD}(x)} \delta_\xi \phi_{\rm BD}(x) \, d^4 x\nonumber\\ &+ \int \frac{\delta S_m}{\delta \psi(x)} \delta_\xi \psi(x) \, d^4 x \, .\nonumber
\end{align}
But now, whereas again the last term is zero on-shell, the second one is not, since the equations of motion for $\phi_{\rm BD}(x)$ are given by:
\begin{equation}
    0 = \int \frac{\delta S_G}{\delta \phi_{\rm BD}(x)} \delta \phi_{\rm BD}(x) \, d^4 x + \int \frac{\delta S_m}{\delta \phi_{\rm BD}(x)} \delta \phi_{\rm BD}(x) \, d^4 x\;,
\end{equation}
[for any variation $\delta \phi_{\rm BD}(x)$, so including $\delta_\xi \phi_{\rm BD}(x)$] and thus if we insist in defining the SET as usual we get:
\begin{equation}
    \nabla^\mu T_{\mu \nu} = \frac{1}{\sqrt{-g}}\frac{\delta S_M}{\delta \phi_{\rm BD}(x)}\nabla_\nu\phi_{\rm BD}(x)\;,
\end{equation}
featuring an apparent non-conservation, as in Rastall's theory.

There are gravitational theories in which the split \eqref{splitGM} is simply impossible. This is the case  presented in Ref.~\cite{Bertini:2019xws}, in which the action is a scalar but the above split is not possible (see also Refs.~\cite{Majerotto:2004ji, Kim:2015iba} for some examples, among several others, within scalar-tensor theories). In these cases, $T_{\mu \nu}$ is commonly defined as a tensor with some properties that resemble, or extend, properties that can be found in the context of GR, although it is not conserved. Whenever the action is a scalar, we know, from the Noether theorem, that there must be a corresponding conserved current, even for the theories in which the split \eqref{splitGM} is impossible; but such conserved current needs not to match the definition in Eq.~\eqref{Tdeff}.

Besides the issue on the split between the gravitational and the matter parts, a scalar action may not lead to a conserved energy-momentum tensor if it depends on external fields. If a field is external, the action variation with respect to it should not be considered: the action only captures part of the fields dynamics, therefore the equations $\delta S_m / \delta \psi  = 0$ do not hold completely, and hence energy-momentum conservation needs not to hold. This case was detailed in Ref.~\cite{Chauvineau:2015cha}.

This results of this section can be directly applied on the specific action-based theories here considered, like $f(R,T)$ and $f(R, {\cal L}_m)$. However, caution is necessary on the $T\mn$ definition that is used. Indeed, for a general $f(R, T)$ theory, there is no reason for the $T\mn$ that appears inside $f(R, T)$ to conserve. Moreover, in general, there is no conserved SET associated to the matter contribution, since there is no general way to decompose such action into a gravitational plus matter parts, as discussed in this appendix.  On the other hand, for any action of the form $f(R, T) = f_1(R) + f_2(T)$ it is possible to define a standard SET that is conserved, since the $f_2(T)$ part can be identified as the matter contribution. Nonetheless, the SET that is conserved is not the same $T\mn$ that appears inside $f_2(T)$. Analogous arguments hold for theories of the type $f(R, {\cal L}_m)$.


\section{From Rastall's stress-energy tensor to an action} \label{sec:TRtoS}

This appendix shows a particular example on how to start from a SET within the Rastall theory, the one satisfies Eq.~\eqref{eq: Rastall conservation}, and to find the corresponding action. That is, we illustrate here the passages:
\begin{equation}
	T\mn^\R \to T\mn^\RC \to  T\mn^\S \to S  \,.
\end{equation} 
Although there is no action whose metric variation will directly lead to $T\mn^\R$, there is a sequence of steps that can always be done to find $T\mn^\R$ from $S$, as commented in Sec.~\ref{Introduction}. The opposite path is not always so clear, since in general it may require field redefinitions.

Here we consider the SET of a free scalar field $\phi$ within Rastall theory. This theory is commonly employed in this sense: one starts from a well-known SET for some type of matter and derive the dynamical consequences of modifying this picture from Rastall's approach, thus implementing Eq.~\eqref{eq: Rastall conservation}. More precisely,  for this scalar field illustration, let 
\begin{equation}
  T^\R\mn = \partial_\mu \phi \partial_\nu \phi - \frac 12 g\mn \partial_\lambda \phi \partial^\lambda \phi \, ,
\end{equation}
where Eq.~\eqref{eq: Rastall conservation} is assumed to hold.

The conserved $T\mn^\RC$ SET is found from Eq.~\eqref{TcalT}, hence:
\begin{equation}
  T\mn^\RC = \partial_\mu  \phi \partial_\nu \phi - \frac 12 k g\mn \partial_\lambda \phi \partial^\lambda \phi \, ,
\end{equation}
with $k \equiv 2- \gamma$. As expected, for $\gamma = 1$, one finds $T\mn^\R = T\mn^\RC$.

We now consider a field redefinition. For clarity, let $k > 0$ and $\partial_\mu \xi\partial^\mu \xi > 0$ (i.e., time-like). If the previous case is not satisfied, one needs to change certain signs properly, but this illustration follows the same steps. Let
\begin{equation}
  \partial_\mu \phi = \frac{1}{\sqrt{k}} (\partial_\nu  \xi \, \partial^\nu \xi)^{\frac{1-k}{2k}} \partial_\mu  \xi \, .
\end{equation}
This redefinition is similar to the one used to interpret a scalar field theory as a perfect fluid (e.g., see Ref.~\cite{Faraoni:2012hn}).

Therefore,
\begin{equation}
  T\mn^\RC = \frac{1}{k} (\partial_\lambda \xi \, \partial^\lambda \xi)^{\frac{1-k}{k}} \left ( \partial_\mu  \xi \partial_\nu \xi - \frac 12 k g_{ab} \partial_\sigma \xi \partial^\sigma \xi \right) .
\end{equation}

This field redefinition is useful since there is an action $S[\xi]$ that leads to $T\mn^\S$, which in turn is identical to $T\mn^\RC$. That is, $T^\RC(\phi) = T^\S(\xi)$. Indeed, the action is 
\begin{equation}
  S[\xi] = \int (\partial_\mu  \xi \partial^\mu  \xi)^{1/{k}} \sqrt{-g} d^4x\, .
\end{equation} 
Although we started from the Rastall picture with the SET of a free scalar field, we see that the same dynamics can be found from GR apart from a field redefinition and a change on the matter content from a free scalar field to a k-essence one. A particular case on the correspondence between the Rastall picture and k-essence appeared previously in Ref.~\cite{Bronnikov:2017pmz}.  The result here presented both illustrates how to find $S$ and generalizes the previous correspondence.

\bibliographystyle{unsrturl}
\bibliography{biblio.bib}

\begin{thebibliography}{10}

\bibitem{Wald:1984rg}
Robert~M. Wald.
\newblock {\em {General Relativity}}.
\newblock Chicago Univ. Pr., Chicago, USA, 1984.
\newblock \href {https://doi.org/10.7208/chicago/9780226870373.001.0001}
  {\path{doi:10.7208/chicago/9780226870373.001.0001}}.

\bibitem{rastall}
P.~Rastall.
\newblock {Generalization of the einstein theory}.
\newblock {\em Phys. Rev. D}, 6:3357--3359, 1972.
\newblock \href {https://doi.org/10.1103/PhysRevD.6.3357}
  {\path{doi:10.1103/PhysRevD.6.3357}}.

\bibitem{Fabris:2011wz}
J.~C. Fabris, M.~Hamani Daouda, and O.~F. Piattella.
\newblock {Note on the Evolution of the Gravitational Potential in Rastall
  Scalar Field Theories}.
\newblock {\em Phys. Lett. B}, 711:232--237, 2012.
\newblock \href {http://arxiv.org/abs/1109.2096} {\path{arXiv:1109.2096}},
  \href {https://doi.org/10.1016/j.physletb.2012.04.020}
  {\path{doi:10.1016/j.physletb.2012.04.020}}.

\bibitem{cosmo}
Carlos E.~M. Batista, Mahamadou~H. Daouda, Julio~C. Fabris, Oliver~F.
  Piattella, and Davi~C. Rodrigues.
\newblock {Rastall Cosmology and the \textbackslash{}Lambda CDM Model}.
\newblock {\em Phys. Rev. D}, 85:084008, 2012.
\newblock \href {http://arxiv.org/abs/1112.4141} {\path{arXiv:1112.4141}},
  \href {https://doi.org/10.1103/PhysRevD.85.084008}
  {\path{doi:10.1103/PhysRevD.85.084008}}.

\bibitem{visser}
Matt Visser.
\newblock {Rastall gravity is equivalent to Einstein gravity}.
\newblock {\em Phys. Lett. B}, 782:83--86, 2018.
\newblock \href {http://arxiv.org/abs/1711.11500} {\path{arXiv:1711.11500}},
  \href {https://doi.org/10.1016/j.physletb.2018.05.028}
  {\path{doi:10.1016/j.physletb.2018.05.028}}.

\bibitem{smalley}
L.~L. {Smalley}.
\newblock {Variational principle for a prototype Rastall theory of
  gravitation}.
\newblock {\em Nuovo Cimento B Serie}, 80:42--48, March 1984.
\newblock \href {https://doi.org/10.1007/BF02899371}
  {\path{doi:10.1007/BF02899371}}.

\bibitem{Sotiriou:2007zu}
Thomas~P Sotiriou, Valerio Faraoni, and Stefano Liberati.
\newblock {Theory of gravitation theories: A No-progress report}.
\newblock {\em Int. J. Mod. Phys. D}, 17:399--423, 2008.
\newblock \href {http://arxiv.org/abs/0707.2748} {\path{arXiv:0707.2748}},
  \href {https://doi.org/10.1142/S0218271808012097}
  {\path{doi:10.1142/S0218271808012097}}.

\bibitem{GabrieleGionti:2020drq}
S.~J. Gabriele~Gionti and S.~J.
\newblock {Canonical analysis of Brans-Dicke theory addresses Hamiltonian
  inequivalence between the Jordan and Einstein frames}.
\newblock {\em Phys. Rev. D}, 103(2):024022, 2021.
\newblock \href {http://arxiv.org/abs/2003.04304} {\path{arXiv:2003.04304}},
  \href {https://doi.org/10.1103/PhysRevD.103.024022}
  {\path{doi:10.1103/PhysRevD.103.024022}}.

\bibitem{Olmo:2011uz}
Gonzalo~J. Olmo.
\newblock {Palatini Approach to Modified Gravity: f(R) Theories and Beyond}.
\newblock {\em Int. J. Mod. Phys. D}, 20:413--462, 2011.
\newblock \href {http://arxiv.org/abs/1101.3864} {\path{arXiv:1101.3864}},
  \href {https://doi.org/10.1142/S0218271811018925}
  {\path{doi:10.1142/S0218271811018925}}.

\bibitem{Toniato:2019rrd}
J\'unior~D. Toniato, Davi~C. Rodrigues, and Aneta Wojnar.
\newblock {Palatini $f(R)$ gravity in the solar system: post-Newtonian
  equations of motion and complete PPN parameters}.
\newblock {\em Phys. Rev. D}, 101(6):064050, 2020.
\newblock \href {http://arxiv.org/abs/1912.12234} {\path{arXiv:1912.12234}},
  \href {https://doi.org/10.1103/PhysRevD.101.064050}
  {\path{doi:10.1103/PhysRevD.101.064050}}.

\bibitem{darabi}
F.~Darabi, H.~Moradpour, I.~Licata, Y.~Heydarzade, and C.~Corda.
\newblock {Einstein and Rastall Theories of Gravitation in Comparison}.
\newblock {\em Eur. Phys. J. C}, 78:25, 2018.
\newblock \href {http://arxiv.org/abs/1712.09307} {\path{arXiv:1712.09307}},
  \href {https://doi.org/10.1140/epjc/s10052-017-5502-5}
  {\path{doi:10.1140/epjc/s10052-017-5502-5}}.

\bibitem{Velten:2021xxw}
Hermano Velten and Thiago R.~P. Caram\^es.
\newblock {To conserve, or not to conserve: A review of nonconservative
  theories of gravity}.
\newblock {\em Universe}, 7(2):38, 2021.
\newblock \href {http://arxiv.org/abs/2102.03457} {\path{arXiv:2102.03457}},
  \href {https://doi.org/10.3390/universe7020038}
  {\path{doi:10.3390/universe7020038}}.

\bibitem{Vanzella:2022xds}
Daniel A.~T. Vanzella.
\newblock {Gravity theories with local energy-momentum exchange: a closer look
  at Rastall's theory}.
\newblock 9 2022.
\newblock \href {http://arxiv.org/abs/2209.01186} {\path{arXiv:2209.01186}}.

\bibitem{harko1}
Tiberiu Harko and Francisco S.~N. Lobo.
\newblock {f(R,$L_{m}$) gravity}.
\newblock {\em Eur. Phys. J. C}, 70:373--379, 2010.
\newblock \href {http://arxiv.org/abs/1008.4193} {\path{arXiv:1008.4193}},
  \href {https://doi.org/10.1140/epjc/s10052-010-1467-3}
  {\path{doi:10.1140/epjc/s10052-010-1467-3}}.

\bibitem{harko2}
Tiberiu Harko, Francisco S.~N. Lobo, Shin'ichi Nojiri, and Sergei~D. Odintsov.
\newblock {$f(R,T)$ gravity}.
\newblock {\em Phys. Rev. D}, 84:024020, 2011.
\newblock \href {http://arxiv.org/abs/1104.2669} {\path{arXiv:1104.2669}},
  \href {https://doi.org/10.1103/PhysRevD.84.024020}
  {\path{doi:10.1103/PhysRevD.84.024020}}.

\bibitem{brasil}
W.~A.~G. De~Moraes and A.~F. Santos.
\newblock {Lagrangian formalism for Rastall theory of gravity and G\"odel-type
  universe}.
\newblock {\em Gen. Rel. Grav.}, 51(12):167, 2019.
\newblock \href {http://arxiv.org/abs/1912.06471} {\path{arXiv:1912.06471}},
  \href {https://doi.org/10.1007/s10714-019-2652-9}
  {\path{doi:10.1007/s10714-019-2652-9}}.

\bibitem{nogales}
Renato Vieira~dos Santos and Jos\'e A.~C. Nogales.
\newblock {Cosmology from a Lagrangian formulation for Rastall's theory}.
\newblock 1 2017.
\newblock \href {http://arxiv.org/abs/1701.08203} {\path{arXiv:1701.08203}}.

\bibitem{Ziaie}
Hamid Shabani and Amir Hadi~Ziaie.
\newblock {A connection between Rastall-type and $f(R, T)$ gravities}.
\newblock {\em EPL}, 129(2):20004, 2020.
\newblock \href {http://arxiv.org/abs/2003.02064} {\path{arXiv:2003.02064}},
  \href {https://doi.org/10.1209/0295-5075/129/20004}
  {\path{doi:10.1209/0295-5075/129/20004}}.

\bibitem{f(R)}
Hans~A. Buchdahl.
\newblock {Non-linear Lagrangians and cosmological theory}.
\newblock {\em Mon. Not. Roy. Astron. Soc.}, 150:1, 1970.

\bibitem{Bertolami}
Orfeu Bertolami, Christian~G. Boehmer, Tiberiu Harko, and Francisco S.~N. Lobo.
\newblock {Extra force in f(R) modified theories of gravity}.
\newblock {\em Phys. Rev. D}, 75:104016, 2007.
\newblock \href {http://arxiv.org/abs/0704.1733} {\path{arXiv:0704.1733}},
  \href {https://doi.org/10.1103/PhysRevD.75.104016}
  {\path{doi:10.1103/PhysRevD.75.104016}}.

\bibitem{Lambdanotconstant}
J.~M. Overduin and F.~I. Cooperstock.
\newblock {Evolution of the scale factor with a variable cosmological term}.
\newblock {\em Phys. Rev. D}, 58:043506, 1998.
\newblock \href {http://arxiv.org/abs/astro-ph/9805260}
  {\path{arXiv:astro-ph/9805260}}, \href
  {https://doi.org/10.1103/PhysRevD.58.043506}
  {\path{doi:10.1103/PhysRevD.58.043506}}.

\bibitem{Lambda(T)}
Nikodem~J. Poplawski.
\newblock {A Lagrangian description of interacting dark energy}.
\newblock 8 2006.
\newblock \href {http://arxiv.org/abs/gr-qc/0608031}
  {\path{arXiv:gr-qc/0608031}}.

\bibitem{Haghani:2023uad}
Zahra Haghani, Tiberiu Harko, and Shahab Shahidi.
\newblock {The first variation of the matter energy-momentum tensor with
  respect to the metric, and its implications on modified gravity theories}.
\newblock 1 2023.
\newblock \href {http://arxiv.org/abs/2301.12133} {\path{arXiv:2301.12133}}.

\bibitem{ma}
Chung-Pei Ma and Edmund Bertschinger.
\newblock {Cosmological perturbation theory in the synchronous and conformal
  Newtonian gauges}.
\newblock {\em Astrophys. J.}, 455:7--25, 1995.
\newblock \href {http://arxiv.org/abs/astro-ph/9506072}
  {\path{arXiv:astro-ph/9506072}}, \href {https://doi.org/10.1086/176550}
  {\path{doi:10.1086/176550}}.

\bibitem{Bertini:2019xws}
Nicolas~R. Bertini, Wiliam~S. Hip\'olito-Ricaldi, Felipe de~Melo-Santos, and
  Davi~C. Rodrigues.
\newblock {Cosmological framework for renormalization group extended gravity at
  the action level}.
\newblock {\em Eur. Phys. J. C}, 80(5):479, 2020.
\newblock [Erratum: Eur.Phys.J.C 80, 644 (2020)].
\newblock \href {http://arxiv.org/abs/1908.03960} {\path{arXiv:1908.03960}},
  \href {https://doi.org/10.1140/epjc/s10052-020-8041-4}
  {\path{doi:10.1140/epjc/s10052-020-8041-4}}.

\bibitem{Majerotto:2004ji}
Elisabetta Majerotto, Domenico Sapone, and Luca Amendola.
\newblock {Supernovae Type Ia data favour negatively coupled phantom energy}.
\newblock 10 2004.
\newblock \href {http://arxiv.org/abs/astro-ph/0410543}
  {\path{arXiv:astro-ph/0410543}}.

\bibitem{Kim:2015iba}
Jik-Su Kim, Chol-Jun Kim, Sin~Chol Hwang, and Yong~Hae Ko.
\newblock {Scalar - Tensor gravity with scalar -matter direct coupling and its
  cosmological probe}.
\newblock {\em Phys. Rev. D}, 96(4):043507, 2017.
\newblock \href {http://arxiv.org/abs/1506.09121} {\path{arXiv:1506.09121}},
  \href {https://doi.org/10.1103/PhysRevD.96.043507}
  {\path{doi:10.1103/PhysRevD.96.043507}}.

\bibitem{Chauvineau:2015cha}
Bertrand Chauvineau, Davi~C. Rodrigues, and J\'ulio~C. Fabris.
\newblock {Scalar\textendash{}tensor theories with an external scalar}.
\newblock {\em Gen. Rel. Grav.}, 48(6):80, 2016.
\newblock \href {http://arxiv.org/abs/1503.07581} {\path{arXiv:1503.07581}},
  \href {https://doi.org/10.1007/s10714-016-2075-9}
  {\path{doi:10.1007/s10714-016-2075-9}}.

\bibitem{Faraoni:2012hn}
Valerio Faraoni.
\newblock {The correspondence between a scalar field and an effective perfect
  fluid}.
\newblock {\em Phys. Rev. D}, 85:024040, 2012.
\newblock \href {http://arxiv.org/abs/1201.1448} {\path{arXiv:1201.1448}},
  \href {https://doi.org/10.1103/PhysRevD.85.024040}
  {\path{doi:10.1103/PhysRevD.85.024040}}.

\bibitem{Bronnikov:2017pmz}
Kirill~A. Bronnikov, J\'ulio~C. Fabris, Oliver~F. Piattella, Denis~C.
  Rodrigues, and Edison~C. Santos.
\newblock {Duality between k-essence and Rastall gravity}.
\newblock {\em Eur. Phys. J. C}, 77(6):409, 2017.
\newblock \href {http://arxiv.org/abs/1701.06662} {\path{arXiv:1701.06662}},
  \href {https://doi.org/10.1140/epjc/s10052-017-4977-4}
  {\path{doi:10.1140/epjc/s10052-017-4977-4}}.

\end{thebibliography}

\end{document}